\documentclass[12pt]{article}

\usepackage{amsmath,amsthm,amsfonts,amssymb,cite}

\let\cal\mathcal

\def\nc{\newcommand}

\def\rom{\rm}
\def\R{\mathbb{R}}
\def\Det{\operatorname{Det}}
\def\Tr{\operatorname{Tr}}
\def\eff{\operatorname{eff}}

\def\const{\operatorname{const}}
\def\cond{\operatorname{cond}}
\def\free{\operatorname{free}}
\def\sign{\operatorname{sign}}
\def\disty{\displaystyle}
\def\I{{\rm i}}
\def\E{{\rm e}}
\def\D{{\rm d}}
\nc{\slacs}[1]{\setlength{\arraycolsep}{#1}}
\nc{\sfrac}[2]{\mbox{$\frac{#1}{#2}$}}
\nc{\bmth}[1]{\mbox{\boldmath$#1$}}

\begin{document}

\title{{\bf The Functional Integration and }\\
   {\bf the Two-Point Correlation Functions } \\
   {\bf  of the Trapped Bose Gas}
}

\author{
$${}$$
$${}$$
\\
{\bf C.~Malyshev,     
     N.~M.~Bogoliubov 
          }
\\
$${}$$
\\
{\it V. A. Steklov Institute of Mathematics,}\\
{\it St.-Petersburg Department,}\\
{\it Fontanka 27, St.-Petersburg, 191023, RUSSIA}\\
$${}$$
}

\date{}

\maketitle

\vskip1.0cm

\begin{abstract}
\noindent
A quantum field-theoretical model, which describes spatially
non-homogeneous repulsive Bose gas in an external harmonic
potential is considered. Two-point thermal correlation functions
of the Bose gas are calculated in the framework of the functional
integration approach. Successive integration over the
``high-energy'' functional variables first and then over the
``low-energy'' ones is used. The effective action functional for
the low-energy variables is obtained in one loop approximation.
The functional integral representations for the correlation
functions are estimated by means of the stationary phase
approximation. A power-law asymptotical behaviour of the
correlators of the one-dimensional Bose gas is demonstrated in the
limit, when the temperature is going to zero, while the volume
occupied by the non-homogeneous Bose gas infinitely increases. The
power-law behaviour is governed by the critical exponent dependent
on the spatial arguments.
\end{abstract}

\newpage

\section{Introduction}

Experimental realization of Bose condensation in vapours of alcali
metals confined in the magneto-optical traps stimulated a
considerable interest to the theory of the Bose gas \cite{pit, 5}.
In particular, Bose condensation in the systems, which are
effectively two-dimensional or quasi one-dimensional, became a
subject of experimental and theoretical investigations. For more
details one should be referred to \cite{pit, 5}. The field models,
which describe the Bose particles with delta-like interparticle
coupling confined by an external harmonic potential, provide a
reliable background for theoretical description of experimental
situations \cite{pit, 5}. For a translationally invariant case,
the field models in question correspond to a quantum nonlinear
Schr\"odinger equation which allows to obtain closed expressions
for the correlation functions in the one-dimensional case
\cite{6}.

Some of the results of the papers \cite{8,9,10} devoted to the
correlation functions of the weakly repulsive Bose gas confined by
an external harmonic potential are reported below. Since there are
no exact solutions in the case of an external potential, the
functional integration approach (see \cite{dev,12,13,14,16,yr,15}
as a list, though incomplete, of appropriate refs.) is used in
\cite{8,9,10} for investigation of the two-point thermal
correlation functions. It will be demonstrated below that the
presence of the external potential results in a modification of
the asymptotical behaviour of the correlation functions in
comparison to a translationally invariant case.

The paper is organized as follows. Section 1 has an introductory
character. A description of the one-dimensional model of
non-relativistic Bose field in question, as well as a summary of
the functional integration approach, are given in Section 2.
Successive integration over the high-energy over-condensate
excitations first and then over the variables, which correspond to
the low-energy quasi-particles, is used in the given paper for a
derivation of one loop effective action for the low excited
quasi-particles. The method of stationary phase is used in Section
3 for approximate investigation of the functional integrals, which
express the two-point thermal correlation functions. Specifically,
the asymptotical approach to estimation of the correlators, which
is discussed in the present paper, was proposed in \cite{17}. It
is clear after \cite{8,9,10} that the method \cite{17} admits a
generalization for the spatially non-homogeneous Bose gas in the
external potential as well. The asymptotics of the two-point
correlation functions of the non-homogeneous one-dimensional Bose
gas are obtained in Section 4. A short discussion in Section 5
concludes the paper.

\section{The effective action and the Thomas\bmth{-}Fermi approximation}

\subsection{The partition function}

Let us consider one-dimensional repulsive Bose gas on the real
axis $\R\ni x$ confined by an external potential $V(x)$. We
represent its partition function $Z$ in the form of the functional
integral \cite{dev,12,13,14,16,yr,15}:
\begin{equation}
\label{1} Z=\int\E^{S[\psi,\bar\psi]}{\cal D}\psi{\cal
D}\bar\psi\,,
\end{equation}
where $S[\psi,\bar\psi]$ is the action functional:
\begin{equation}
\label{2}
\begin{array}{rcl}
S[\psi,\bar\psi]&=&\disty\int_0^\beta\D\tau\int\D x
\Big\{\bar\psi(x,\tau)\Big(\frac\partial{\partial\tau}- {\cal
H}\Big)\psi(x,\tau)-\\[6pt]
&&\disty\qquad\qquad-\frac{g}2\,\bar\psi(x,\tau)\bar\psi(x,\tau)\psi(x,\tau)
\psi(x,\tau)\Big\}\,,
\end{array}
\end{equation}
and ${\cal D}\psi{\cal D}\bar\psi $ is the functional integration
measure. Other notations in (\ref{1}), (\ref{2}) are: ${\cal H}$
is the ``single-particle'' Hamiltonian,
\begin{equation}
\label{3}
\mathcal{H}\equiv\frac{-\hbar^2}{2m}\,\frac{\partial^2}{\partial
x^2}-\mu+V(x)\,,
\end{equation}
$m$ is the mass of the Bose particles, $\mu$ is the chemical
potential, $g$ is the coupling constant corresponding to the weak
repulsion (i.e., $g>0$), and the external confining potential is
$V(x)\equiv\frac12\,m\Omega^2x^2$. The domain of the functional
integration in (\ref{1}) is given by the space of the
complex-valued functions $\bar\psi(x,\tau)$, $\psi(x,\tau)$
depending on $x\in\R$ and $\tau\in[0,\beta]$. With regard to $x$,
the functions $\bar\psi(x,\tau)$, $\psi(x,\tau)$ belong to the
space of quadratically integrable functions $L_2(\R)$, while they
are finite and periodic with the period $\beta =(k_BT)^{-1}$ with
regard to the imaginary time $\tau$ ($k_B$ is the Boltzmann
constant, and $T$ is an absolute temperature).

At sufficiently low temperatures each of the variables $\bar\psi (x, \tau)$,
$\psi (x, \tau)$ is given by two constituents:
\begin{equation}
\psi(x,\tau)=\psi_o(x,\tau)+\psi_e(x,\tau)\,,\quad
\bar\psi(x,\tau)=\bar\psi_o(x,\tau)+\bar\psi_e(x,\tau)\,,
\label{4}
\end{equation}
where $\bar\psi_o(x,\tau)$, $\psi_o(x,\tau)$ correspond to
\textit{quasi-condensate} (a true Bose condensate does not exist
in one-dimensional systems \cite{16}), while $\bar\psi_e(x,\tau)$,
$\psi_e(x,\tau)$ correspond to the high-energy thermal (i.e.,
over-condensate) excitations. In the exactly solvable case, the
existence of the quasi-condensate implies that a non-trivial
ground state exists \cite{6}. Let us require the variables
(\ref{4}) to be orthogonal in the following sense:
$$
\int\psi_o(x,\tau)\bar\psi_e(x,\tau)\,\D x=
\int\bar\psi_o(x,\tau)\psi_e(x,\tau)\,\D x=0\,.
$$
Then, the integration measure ${\cal D}\psi{\cal D}\bar\psi$ is
replaced by the measure ${\cal D}\psi_o{\cal D}\bar\psi_o{\cal
D}\psi_e{\cal D}\bar\psi_e$.

To investigate the functional integral (\ref{1}), we shall perform
a successive integration: first, we shall integrate over the
high-energy constituents given by (4), and then over the
low-energy ones \cite{12, 16}. At a second step, it is preferable
to pass to new functional variables, which describe an observable
``low-energy'' physics in a more adequate way. After the
substitution of (\ref{4}) into the action (\ref{2}) we take into
account in $S$ only the terms up to quadratic in $\bar\psi_e$,
$\psi_e$. This means an approximation, in which the
over-condensate quasi-particles do not couple with each other. In
this case, it is possible to integrate out the thermal
fluctuations $\bar\psi_e$, $\psi_e$ in a closed form and thus to
arrive to an effective action functional $S_{\eff}[\psi_o, \bar
\psi_o]$. It depends only on the quasi-condensate variables
$\psi_o, \bar \psi_o$:
\begin{equation}
S_{\eff}[\psi_o,\bar\psi_o]=\ln\int\E^{\widetilde
S[\psi_o+\psi_e,\bar\psi_o+\bar\psi_e]}{\cal D}\psi_e{\cal D}\bar
\psi_e\,,
\label{5}
\end{equation}
where the tilde in $\widetilde S$ implies that ``self-coupling''
of the fields $\bar\psi_e$, $\psi_e$ is excluded. With respect of
(\ref{5}), the partition function of the model $Z$ (\ref{1}) takes
an approximate form:
\begin{equation}
Z \approx \int \E^{S_{\eff}[\psi_o,\bar \psi_o]}{\cal
D}\psi_o{\cal D}\bar\psi_o\,.
\label{6}
\end{equation}

Let us consider the derivation of the effective action
$S_{\eff}[\psi_o,\bar\psi_o]$ (\ref{5}) in more details. The
splitting (\ref{4}) allows to derive $S_{\eff}[\psi_o,\bar\psi_o]$
in the framework of the field-theoretical approach of loop
expansion \cite{itz}. We substitute (\ref{4}) into the initial
action $S[\psi,\bar\psi]$ (\ref{2}) and then go over from $S$ to
the action $\widetilde S$, which is given by three terms:
\begin{equation}
\widetilde{S}=S_{\cond}+S_{\free}+S_{\mathrm{int}}\,.
\label{7}
\end{equation}
In (\ref{7}), $S_{\cond}$ is the action functional of the
condensate quasi-particles, which corresponds to a tree
approximation \cite{itz}:
\begin{equation}
\begin{array}{rcl}
S_{\cond}[\psi_o,\bar\psi_o]&\equiv&\disty \int_0^\beta\D\tau
\int\D x\Big\{\bar\psi_o(x,\tau)\widehat K_{+}\psi_o(x,\tau)-\\[9pt]
&&\disty\qquad\quad
-\frac{g}2\,\bar\psi_o(x,\tau)\bar\psi_o(x,\tau)\psi_o(x,\tau)\psi_o(x,\tau)\Big\}.
\end{array}
\label{8}
\end{equation}
At the chosen approximation the action for the over-condensate
excitations $S_{\free}$ takes the form:
\begin{equation}
S_{\free}[\psi_e,\bar\psi_e]\equiv \frac12\int_0^\beta\D\tau
\int\D x\bigl(\bar\psi_e,\psi_e\bigr)\widehat G^{-1}
\begin{pmatrix}\psi_e \\
                \bar \psi_e
\end{pmatrix}\,,
\label{9}
\end{equation}
where $\widehat{G}^{-1}$ is the matrix-differential operator,
\begin{equation}
\widehat{G}^{-1}\equiv\widehat{G}_0^{-1}-\widehat{\Sigma}\,.
\label{10}
\end{equation}
In (\ref{10}) we defined:
\begin{equation}
{\slacs{1.5ex}
\widehat{G}_0^{-1}\equiv\begin{pmatrix}
                  \widehat{K}_{+} & 0 \\
                  0 & \widehat{K}_{-}
\end{pmatrix},
\qquad \widehat{\Sigma}\equiv\widehat{\Sigma}[\psi_o,\bar\psi_o]=g
\begin{pmatrix}
                2\bar\psi_o \psi _o & \psi_o^2 \\
                (\bar \psi_o)^2 & 2\bar\psi_o\psi_o
\end{pmatrix},}
\label{11}
\end{equation}
where $\widehat{K}_{\pm}$ are the differential operators,
$\widehat{K}_{\pm}\equiv\pm\dfrac{\partial}{\partial\tau}-\mathcal{H}$,
and ${\cal H}$ is the single-particle Hamiltonian (\ref{3}).
Eventually, $S_{\mathrm{int}}$ describes a coupling of the
quasi-con\-den\-sate to the over-condensate excitations:
\begin{equation}
\begin{array}{rcl}
\disty S_{\rom{int}}[\psi_o,\bar\psi_o,\psi_e,\bar\psi_e]&\equiv&
\disty\int_0^\beta\D\tau\int\D x
\Big\{\bar\psi_e(x,\tau)\bigl[\widehat{K}_{+}-
                      g\bar\psi_o\psi_o\bigr]\psi_o(x,\tau)+\\[9pt]
&&\disty\qquad\quad
+\psi_e(x,\tau)\bigl[\widehat{K}_{-}-g\bar\psi_o\psi_o\bigr]\bar\psi_o(x,\tau)\Big\}.
\end{array}
\label{12}
\end{equation}

It is appropriate to apply the stationary phase method to the
functional integral (\ref{6}). To this end, let us choose
$\bar\psi_o$, $\psi_o$ as the stationarity points of the
functional $S_{\cond}$ (\ref{8}), which are defined by the
extremum condition $\delta(S_{\cond}[\psi_o,\bar\psi_o])=0$. The
corresponding equations look like the Gross--Pitaevskii-type
equations \cite{pit}:
\begin{equation}
\begin{array}{rcl}
\biggl(\dfrac\partial{\partial\tau}+\dfrac{\hbar^2}{2m}\,
\dfrac{\partial^2}{\partial x^2}+\mu-V(x)\biggr)\psi_o-
g\left(\bar\psi_o\psi_o\right)\psi_o&=&0\,,\\[9pt]
\biggl(-\dfrac\partial{\partial\tau}+\dfrac{\hbar^2}{2m}\,
\dfrac{\partial^2}{\partial x^2}+\mu-V(x)\biggr)\bar\psi_o-
g\left(\bar\psi_o\psi_o\right)\bar\psi_o&=&0\,.
\end{array}
\label{13}
\end{equation}
The contribution of the action functional $S_{\mathrm{int}}$
(\ref{12}) drops out from (\ref{7}), provided $\bar\psi_o$,
$\psi_o$ are solutions of equations (\ref{13}). Therefore the
dynamics of $\psi_e$, $\bar\psi_e$ is described, in the leading
approximation, by the action $S_{\free}$ (\ref{9}). The latter
depends on $\bar\psi_o$, $\psi_o$ non-trivially through the matrix
of the self-energy parts $\widehat{\Sigma}$, which enters into
$\widehat{G}^{-1}$ (\ref{10}).

The \textit{Thomas--Fermi approximation} is essentially used in
the present paper in order to determine the stationarity points
$\bar\psi_o$, $\psi_o$. This approximation consists in neglect of
the kinetic term $\dfrac{\hbar^2}{2m}\,\dfrac{\partial^2}
{\partial x^2}$ in equations (\ref{13}) \cite{pit, 5}. The
Thomas--Fermi approximation is valid for the systems containing a
sufficiently large number of particles in the magneto-optical
traps \cite{pit, 5}. The following condensate $\tau$-independent
solution can be obtained:
\begin{equation}
\bar\psi_o\psi_o=\rho_{\mathrm{TF}}(x;\mu)\equiv
\frac1g\,\bigl(\mu-V(x)\bigr) \Theta\bigl( \mu -V(x)\bigr)\,,
\label{14}
\end{equation}
where $\Theta$ is the Heavyside function. Now the integration in
(\ref{6}) with respect to $\psi_e$, $\bar\psi_e$ is Gaussian. This
leads to the one loop effective action in terms of the variables
$\psi_o$, $\bar\psi_o$:
\begin{equation}
S_{\eff}[\psi_o,\bar\psi_o]\equiv S_{\cond}[\psi_o,\bar\psi_o]-
\sfrac12\,\ln\Det\left(\widehat{G}^{-1}\right).
\label{15}
\end{equation}
Here $\widehat{G}^{-1}$ is the matrix operator (\ref{10}) and
$\psi_o$, $\bar\psi_o$ have a sense of the new variables governed
by the action (\ref{15}).

In order to assign a meaning to the final expression for the
effective action (15), it is necessary to regularize the
determinant $\Det\left(\widehat{G}^{-1}\right)$. In our case, the
operator $\widehat{G}^{-1}$ is already written as
$2\times2$-matrix Dyson equation (\ref{10}), where the entries of
$\widehat{\Sigma }[\psi_o,\bar\psi_o]$ (\ref{11}) play the role of
the normal ($\Sigma_{11}=\Sigma_{22}$) and anomalous
($\Sigma_{12}$, $\Sigma_{21}$) self-energy parts. The Dyson
equation (\ref{10}) defines the matrix $\widehat{G}$, where the
entries have a meaning of the Green functions of the fields
$\bar\psi_e$, $\psi_e$. The matrix $\widehat{G}$ arises as a
formal inverse of the operator $\widehat{G}^{-1}$.

It is appropriate to represent $\widehat{G}^{-1}$ as follows:
\begin{equation}
\widehat{G}^{-1}=\widehat{G}_0^{-1}-\widehat{\Sigma }\equiv
\widehat{\mathcal{G}}^{-1}-\left(\widehat{\Sigma}-2g\rho_{\mathrm{TF}}(x;\mu)
\widehat I\right),
\label{16}
\end{equation}
where $\widehat I$ is the unit matrix of the size $2\times2$, and
the matrix $\widehat{{\cal G}}^{-1}$ is defined as
\begin{equation}
{\slacs{1.5ex}
\widehat{{\cal G}}^{-1}\equiv
\begin{pmatrix}
         \widehat{K}_{+}-2g\rho _{\mathrm{TF}}(x;\mu ) & 0 \\
         0 & \widehat{K}_{-}-2g\rho _{\mathrm{TF}}(x;\mu )
\end{pmatrix}
\equiv
\begin{pmatrix}
              {\cal K}_{+} & 0 \\
              0 & {\cal K_{-}}
\end{pmatrix}.}
\label{17}
\end{equation}
Here $\rho_{\mathrm{TF}}(x;\mu)$ is the solution (\ref{14}) and
equation (\ref{16}) implies that we simply added and subtracted
$2g\rho_{\mathrm{TF}}(x;\mu)$ on the principle diagonal of the
matrix operator $\widehat{G}^{-1}$. A formal inverse of
$\widehat{{\cal G}}^{-1}$ can be found from the following
equation, which defines the Green functions ${\cal G}_{\pm}$:
$$
{\slacs{1.5ex}
\begin{pmatrix}
           {\cal K}_{+} & 0 \\
           0 & {\cal K_{-}}
\end{pmatrix}
\begin{pmatrix}
          {\cal G}_{+} & 0 \\
          0 & {\cal G}_{-}
\end{pmatrix}
=\delta(x-x^{\prime})\delta(\tau-\tau^{\prime})\widehat{I}\,.}
$$

Using the relation $\ln\,\Det=\Tr\,\ln$, one gets:
\begin{equation}
{\slacs{1ex} -\frac12\ln\,\Det\left(\widehat{G}^{-1}\right)=
-\frac12\Tr\,\ln\left(\widehat
I-\widehat{\mathcal{G}}\bigl(\widehat{\Sigma}-2g\rho_{\mathrm{TF}}(x;\mu)
\widehat I\bigr)\right)-\frac12\,\ln\,\Det
\begin{pmatrix}
            {\cal K}_{+} & 0 \\
            0 & {\cal K_{-}}
\end{pmatrix}.}
\label{18}
\end{equation}
The first term in right-hand side of (\ref{18}) is free from
divergencies. Let us consider the determinant of the
matrix-differential operator in right-hand side of (\ref{18}). Let
us denote the eigenvalues of the operators ${\cal K}_{\pm }$ as
$\pm\I\omega_B-\lambda_n$, where $\omega_B$ are the bosonic
Matsubara frequencies and $\lambda_n$ are the energy levels
labeled by the multi-index $n$ \cite{9,10}. Then, we calculate
\cite{9,10}:
$$
{\slacs{1ex}
\frac1{2\beta}\,\ln\Det
\begin{pmatrix}
           {\cal K}_{+} & 0 \\
           0 & {\cal K_{-}}
\end{pmatrix}=\frac1\beta\sum_n\ln\left(2\sinh\frac{\beta\lambda_n}2\right)\equiv
\widetilde{F}_{\mathrm{nc}}(\mu )\,,}
$$
where $\widetilde{F}_{\mathrm{nc}}$ has a sense of the free energy
of an ideal gas of the over-condensate excitations. Then, in the
leading order in $g$, one gets:
\begin{eqnarray}
&&-\frac12\ln\Det\left(\widehat{G}^{-1}\right)\approx
-\beta\widetilde{F}_{\mathrm{nc}}(\mu)+\nonumber\\
&&\qquad\qquad+g\int_0^\beta\D\tau \int\D
x\bigl(\mathcal{G}_{+}(x,\tau;x,\tau)+\mathcal{G}_{-}(x,\tau;x,\tau)\bigr)
\bigl(\bar\psi_o\psi_o-\rho_{\mathrm{TF}}(x;\mu)\bigr)\equiv\nonumber\\
&&\qquad\equiv -\beta F_{\mathrm{nc}}(\mu )-2g\int_0^\beta
\D\tau \int \D x\,\rho_{\mathrm{nc}}(x)\bar\psi_o\psi_o\,. \label{19}
\end{eqnarray}
\unskip Here $F_{\mathrm{nc}}$ is the free energy of the non-ideal
gas of the over-condensate quasi-particles. The density of the
over-condensate quasi-particles is
$\rho_{\mathrm{nc}}(x)\equiv-\mathcal{G}_{\pm}(x,\tau; x,\tau)$, and it 
depends only on the spatial coordinate $x$. At
very low temperatures and sufficiently far from the boundary of
the domain occupied by the condensate, the quantity
$\rho_{\mathrm{nc}}(x)$ can approximately be replaced by
$\rho_{\mathrm{nc}}(0)$, since ${\cal G}_{\pm }(x,\tau;x,\tau)$ is
almost constant over a considerable part of the condensate \cite{nar}.

It is appropriate to write the one loop effective action obtained
in terms of new independent real-valued variables of the
functional integration. Namely, in terms of the density
$\rho(x,\tau)$ and the phase $\varphi(x,\tau)$ of the field
$\psi_o(x,\tau)$:
\begin{equation}
\psi_o(x,\tau)=\sqrt{\rho(x,\tau)}\,\E^{\I\varphi(x,\tau)}\,,\qquad
\bar\psi_o(x,\tau)=\sqrt{\rho(x,\tau)}\,\E^{-\I\varphi(x,\tau)}\,.
\label{20}
\end{equation}
In terms of these variables, the effective action takes the form
\cite{9,10}:
\begin{equation}
\begin{array}{l}
\disty S_{\eff}[\rho,\varphi]=-\beta
F_{\mathrm{nc}}(\mu)+\I\int_0^\beta\D\tau \int\D
x\biggl\{\rho\partial_\tau\varphi+\frac{\hbar^2}{2m}\,
\partial_x(\rho\partial_x\varphi)\biggr\}+\\[9pt]
\disty\qquad\qquad+\int_0^\beta\D\tau \int\D
x\biggl\{\frac{\hbar^2}{2m}\,\Bigl(\sqrt{\rho}\,
\partial_x^2\sqrt{\rho}-\rho(\partial_x\varphi)^2\Bigr)+
(\Lambda-V)\rho-\frac{g}2\,\rho^2\biggr\},
\end{array}
\label{21}
\end{equation}
where $\Lambda=\mu-2g\rho_{\mathrm{nc}}(0)$ is the renormalized
chemical potential. Here and below we denote the partial
derivatives of the first order over $\tau$ and~$x$ as
$\partial_\tau$ and $\partial_x$, respectively, whereas the
partial derivatives of the second order --- as $\partial_\tau^2$
and $\partial_x^2$. The model in question in the present paper is
spatially one-dimensional, and a possible multi-valuedness of the
angle variable $\varphi$ is left aside.

We shall consider $S_{\eff}$ (\ref{21}) as the one loop effective
action, where the thermal corrections over the ``classical''
background are taken into account. The ``classical'' background
corresponds to the solution (\ref{14}). It should be noticed that
our derivation of the effective action can formally be used for
two and three dimensions also \cite{8}. Notice that equation
(\ref{21}) remains correct at $V=0$ also.

\subsection{The excitation spectrum}

Let us determine the spectrum of the low-energy quasi-particles.
Now we are applying the stationary phase approximation to the
integral (\ref{6}), where the effective action is given by
(\ref{21}), while the measure is ${\cal D}\rho {\cal D}\varphi$.
The corresponding stationarity point is given by the extremum
condition $\delta \bigl(S_{\eff}[\rho,\varphi]\bigr)=0$, which is
equivalent to the couple of the Gross--Pitaevskii equations:
\begin{equation}
\begin{array}{r}
\disty
\I\partial_\tau\varphi+\frac{\hbar^2}{2m}\left(\frac1{\sqrt{\rho}}\,
\partial_x^2\sqrt{\rho}-(\partial_x\varphi)^2\right)+\Lambda-V(x)-g\rho=0\,,\\[6pt]
-\I\partial_\tau\rho+\dfrac{\hbar^2}m\,
\partial_x\left(\rho\partial_x\varphi\right)=0\,.
\end{array}
\label{22}
\end{equation}
We use the Thomas--Fermi approximation and drop out the term
$(\partial_x^2\sqrt{\rho})/\sqrt{\rho}$ in the first equation in
(\ref{22}). Solution with
$\partial_\tau\rho=0=\partial_\tau\varphi$ appears, provided the
velocity field $\mathbf{v}=m^{-1}\partial_x\varphi$ is taken equal
to zero in (\ref{22}). In this case, equations (\ref{22}) lead to
the density of the condensate:
\begin{equation}
\rho_{\mathrm{TF}}(x)\equiv\frac{\Lambda}{g}\,
\widetilde{\rho}_{\mathrm{TF}}(x)=\frac{\Lambda}g\,\left(1-\frac{x^2}{R_c^2}\right)
\Theta\left(1-\frac{x^2}{R_c^2}\right).
\label{23}
\end{equation}
Explicit form of the external potential
$V(x)=\frac12\,m\Omega^2x^2$ is taken into account in the
expression (\ref{23}). The form of the solution (\ref{23}) means
that the quasi-condensate occupies the domain $|x|\le R_c$ at zero
temperature. The length $R_c$ defines the boundary of this domain,
$R_c^2\equiv\dfrac{2\Lambda}{m\Omega^2}$ (in three dimensional
space, this would correspond to a spherical distribution of the
condensate). In the homogeneous case given by the limit
$1/R_c\to0$, the Thomas--Fermi solution $\rho_{\mathrm{TF}}(x)$ is
transformed into the density $\rho_{\mathrm{TF}}(0)=\Lambda/g$,
which coincides with the density of the homogeneous Bose gas
\cite{12}.

According to the initial splitting (\ref{4}), we suppose that
thermal fluctuations in vicinity of the stationarity point
(\ref{23}) are small, and therefore an analogous splitting can be
written for the condensate density also:
\begin{equation}
\rho_0(x,\tau)=\rho_{\mathrm{TF}}(x)+\pi_0(x,\tau)\,,
\label{24}
\end{equation}
where $\rho_0(x,\tau)$ is a specific solution of (\ref{22}). We
linearize equations (\ref{22}) in a vicinity of the equilibrium
solution $\rho_0=\rho_{\mathrm{TF}}(x)$, $\varphi=\const$.
Eliminating the phase $\varphi$ and dropping out the terms
proportional to $\hbar^4$, we go over from (\ref{22}) to the
\textit{Stringari thermal equation} \cite{30}:
\begin{equation}
\frac1{\hbar^2v^2}\,\partial_\tau^2\pi_0+
\partial_x\biggl(\Big(1-\frac{x^2}{R_c^2}\Big)
\partial_x\pi_0\biggr)=0\,,
\label{25}
\end{equation}
where the parameter $v$ has a meaning of the sound velocity in the
center of the trap:
\begin{equation}
v^2\equiv\frac{\rho_{\mathrm{TF}}(0)g}m=\frac{\Lambda}m\,.
\label{26}
\end{equation}

The substitution $\pi_0=\E^{\I\omega\tau}u(x)$ transforms
(\ref{25}) into the Legendre equation:
\begin{equation}
-\frac{\omega^2}{\hbar^2v^2}\,u(x)+\frac{\D}{\D x}
\biggl(\Big(1-\frac{x^2}{R_c^2}\Big)\frac{\D}{\D
x}u(x)\biggr)=0\,.
\label{27}
\end{equation}
Since the Thomas--Fermi solution (\ref{23}) is non-zero only at
$|x|\le R_c$, we shall consider (\ref{27}) at
$x\in[-R_c,R_c]\subset\R$, as well. After an analytical
continuation $\omega\rightarrow \I E$, equation (\ref{27})
possesses the polynomial solutions, which are given by the
Legendre polynomials $P_n(x/R_c)$, if and only if
\begin{equation}
\left(\frac{R_c}{\hbar v}\right)^2E^2\equiv
\frac2{\hbar^2\Omega^2}\,E^2=n(n+1)\,,\qquad n\geq 0\,.
\label{28}
\end{equation}
In other words, equation (\ref{27}) leads to the spectrum of the
low lying excitations: $E_n=\hbar\Omega\sqrt{\dfrac{n(n+1)}2}$,
$n\ge0$, \cite{31}. Notice that the corresponding equation for the
homogeneous Bose gas is obtained after a formal limit $1/R_c\to 0$
in (\ref{27}) at finite $x$. Provided the latter is still
considered for the segment $[-R_c,R_c]\ni x$ with a periodic
boundary condition for $x$, we arrive at the discrete spectrum of
the following form: $E_k=\hbar v k$, where $k$ is the wave number,
$k=(\pi/R_c)n$, $n\in\mathbb{Z}$.

\section{The two-point correlation functions}

Let us go over to our main task --- to the calculation of the
two-point thermal correlation function
$\Gamma(x_1,\tau_1;x_2,\tau_2)$ of the spatially non-homogeneous
Bose gas. We define it as the ratio of two functional integrals:
\begin{equation}
\Gamma(x_1,\tau_1;x_2,\tau_2)=\frac{\int\bar\psi(x_1,\tau_1)\psi(x_2,\tau_2)
\E^{S[\psi,\bar\psi]}{\cal D}\psi{\cal D}\bar\psi}
{\int\E^{S[\psi,\bar\psi]}{\cal D}\psi{\cal D}\bar\psi}\,,
\label{29}
\end{equation}
where the action $S[\psi,\bar\psi]$ is given by (\ref{2}).

We are interested in the behaviour of the correlators at the
distances considerably smaller in comparison with the size of the
domain occupied by the condensate. The main contribution to the
behaviour of the correlation functions is due to the low lying
excitations at sufficiently low temperatures \cite{12,16}. To
calculate $\Gamma(x_1,\tau_1;x_2,\tau_2)$, (\ref{29}), we use the
method of successive functional integration first over the
high-energy excitations $\bar\psi_e$, $\psi_e$, and then over the
low-energy excitations $\bar\psi_o$, $\psi_o$ (see (\ref{4})). In
the leading approximation, the correlator we are interested in
looks, in terms of the density--phase variables, as follows
\cite{9,10}:
\begin{equation}
\begin{array}{rcl}
\Gamma (x_1,\tau_1;x_2,\tau_2)&\simeq&\disty
\int\exp\Bigl(S_{\eff}[\rho,\varphi]-
\I\varphi(x_1,\tau_1)+\I\varphi(x_2,\tau_2)+\\[6pt]
&&\qquad\qquad\qquad
+\frac12\,\ln\rho(x_1,\tau_1)+\frac12\,\ln\rho(x_2,\tau_2)\Bigr)
{\cal D}\rho {\cal D}\varphi\times\\[6pt]
&&\disty\times\left(\int\exp\bigl(S_{\eff}[\rho,\varphi]\bigr)
{\cal D}\rho{\cal D}\varphi\right)^{-1},
\end{array}
\label{30}
\end{equation}
where the integrand in the nominator is arranged in the form of a
single exponential. Here $S_{\eff}[\rho,\varphi]$ is the effective
action (\ref{21}).

Since the fluctuations of the density are suppressed at
sufficiently low temperatures \cite{nar}, one can replace
$\ln\rho(x_1,\tau_1)$, $\ln\rho(x_2,\tau_2)$ in (\ref{30}) by
$\ln\rho_{\mathrm{TF}}(x_1)$, $\ln\rho_{\mathrm{TF}}(x_2)$, where
$\rho_{\mathrm{TF}}$ is defined by (\ref{23}). In accordance with
the variational principle suggested in \cite{17}, we estimate the
functional integrals in (\ref{30}) by the stationary phase method.
Each of the integrals is characterized by its own stationarity
point given by variation of the corresponding exponent. For the
correlation function $\Gamma(x_1,\tau_1;x_2,\tau_2)$, we obtain
the following leading estimation:
\begin{equation}
\begin{array}{l}
\Gamma(x_1,\tau_1;x_2,\tau_2)\simeq
\sqrt{\rho_{\mathrm{TF}}(x_1)\rho_{\mathrm{TF}}(x_2)}\times\\[7pt]
\qquad\qquad\qquad
\times\exp\bigl(-S_{\eff}[\rho_0,\varphi_0]{+}S_{\eff}[\rho_1,\varphi_1]-
\I\varphi_1(x_1,\tau_1)+\I\varphi_1(x_2,\tau_2)\bigr)\,,
\end{array}
\label{31}
\end{equation}
where the variables $\rho_0$, $\varphi_0$ are defined by the
extremum condition $\delta\bigl(S_{\eff}[\rho,\varphi]\bigr)=0$,
and therefore they just satisfy the Gross--Pitaevskii equations
(\ref{22}). The fields $\rho_1$, $\varphi_1$ are defined by the
extremum condition:
\begin{equation}
\delta\Bigl(S_{\eff}[\rho,\varphi]-\I\varphi(x_1,\tau_1)+
\I\varphi(x_2,\tau_2)\Bigr)=0\,.
\label{32}
\end{equation}

The variational equation (\ref{32}) leads to another couple of
equations of the Gross--Pitaevskii type. One of these equations
turns out to be a non-homogeneous equation with the $\delta$-like
source, while another one is a homogeneous equation. In fact, the
homogeneous equation appears due to a requirement of vanishing of
the coefficient at the variation $\delta\rho(x,\tau)$, while the
non-homogeneous equation is defined by vanishing of the
coefficient at the variation $\delta\varphi(x,\tau)$.

It can consistently be assumed that the solution $\rho_1(x,\tau)$
can be represented as a sum of $\rho_{\mathrm{TF}}(x)$ and of a
weakly fluctuating part, provided the boundary $R_c$ is far from
beginning of coordinates:
$\rho_1(x,\tau)=\rho_{\mathrm{TF}}(x)+\pi_1(x,\tau)$. Therefore,
the terms $\sqrt{\pi_1}\,\partial_x^2\sqrt{\pi_1}$ and
$\partial_x\pi_1\partial_x\varphi_1$ are small and can be omitted.
Taking into account a linearization near the Thomas--Fermi
solution, one can finally arrive at a couple of the following
equations:
\begin{equation}
\begin{array}{rcl}
\I\partial_\tau\varphi_1-g\pi_1-
\dfrac{\hbar^2}{2m}\left(\partial_x\varphi_1\right)^2&=&0\,,\\[6pt]
-\I\partial_\tau\pi_1+\dfrac{\hbar^2}m\,
\partial_x\left(\rho_{\mathrm{TF}}\partial_x\varphi_1\right)&=&
\I\delta(x-x_1)\delta(\tau-\tau_1)-\I\delta(x-x_2)\delta(\tau-\tau_2)\,.
\end{array}
\label{33}
\end{equation}
Equations (\ref{33}) lead \cite{9,10} to the following equation
for the variable $\varphi_1$:
\begin{equation}
\frac1{\hbar^2v^2}\,\partial_\tau^2\varphi_1+
\partial_x\left(\widetilde\rho_{\mathrm{TF}}(x)\partial_x\varphi_1\right)=
\I\,\frac{mg}{\hbar^2\Lambda}\Bigl\{\delta(x-x_1)\delta(\tau-\tau_1)-
\delta(x-x_2)\delta(\tau-\tau_2)\Bigr\},
\label{34}
\end{equation}
where $v$ means the sound velocity in the center of the trap
(\ref{26}), and $\widetilde\rho_{\mathrm{TF}}$ is defined by (23).
Now, with the help of (\ref{33}) one can calculate the terms
contributing into the exponent in (\ref{31}), \cite{9,10}:
\begin{equation}
-S_{\eff}[\rho_0,\varphi_0]+S_{\eff}[\rho_1,\varphi_1]\simeq
\frac{\I}2\,\bigl(\varphi_1(x_1,\tau_1)-\varphi_1(x_2,\tau_2)\bigr)\,.
\label{35}
\end{equation}
Substituting (\ref{35}) into (\ref{31}), one obtains the following
approximate formula for the correlator:
\begin{equation}
\Gamma(x_1,\tau_1;x_2,\tau_2)\simeq
\sqrt{\rho_{\mathrm{TF}}(x_1)\rho_{\mathrm{TF}}(x_2)}\,
\exp\left(-\frac{\I}2\,\bigl(\varphi_1(x_1,\tau_1)-
\varphi_1(x_2,\tau_2)\bigr)\right).
\label{36}
\end{equation}

It is natural to represent solutions of equation (\ref{34}) in
terms of the solution $G(x,\tau;x^{\prime},\tau^{\prime})$ of the
equation
\begin{equation}
\frac1{\hbar^2v^2}\,\partial_\tau^2G(x,\tau;x^\prime,\tau^\prime)+
\partial_x\left(\Bigl(1-\frac{x^2}{R_c^2}\Bigr)\partial_x G(x,\tau;
x^\prime,\tau^\prime)\right)=
\frac{g}{\hbar^2v^2}\,\delta(x-x^\prime)\delta(\tau-\tau^\prime)\,.
\label{37}
\end{equation}
Bearing in mind the homogeneous equation (\ref{25}), we shall call
(\ref{37}) as \textit{non-ho\-mo\-ge\-ne\-ous Stringari equation}.
As it is clear after \cite{pp}, the Green function
$G(x_1,\tau_1;x_2,\tau_2)$ has a meaning of the correlation
function of the phases:
\begin{equation}
G(x_1,\tau_1;x_2,\tau_2)=-\bigl<\varphi(x_1,\tau_1)\varphi(x_2,\tau_2)\bigr>,
\label{38}
\end{equation}
where the angle brackets in right-hand side should be understood
as an averaging with respect to the weighted measure
$\exp{\bigl(S_{\eff}[\rho,\varphi ]\bigr)}\,{\cal D}\rho {\cal
D}\varphi$. Using (\ref{38}), it is possible to represent,
eventually, the correlation function as follows \cite{9,10}:
\begin{equation}
\Gamma(x_1,\tau_1;x_2,\tau_2)\simeq\sqrt{\widetilde\rho(x_1)\widetilde\rho(x_2)}\,
\exp\left(-\sfrac12\,\bigl(G(x_1,\tau_1;x_2,\tau_2)+
G(x_2,\tau_2;x_1,\tau_1)\bigr)\right),
\label{39}
\end{equation}
where $\widetilde\rho(x_1)$, $\widetilde\rho(x_2)$ are the
renormalized densities. The solution $G(x_1,\tau_1;x_2,\tau_2)$ of
equation (\ref{37}) is defined up to a purely imaginary additive
constant, which has a meaning of a global phase.

The governing equations reported in \cite{8} for the spatial
dimensionalities $d=3$, 2, 1 are in a direct agreement with
(\ref{33}), provided the $\tau$-dependence is neglected in
(\ref{33}). The correlation functions of the phases are obtained
in \cite{8} without an influence of
$\partial_x\rho_{\mathrm{TF}}\,\partial_x\varphi_1$ as follows
(the notation $f(\mathbf{x},\mathbf{x}^\prime)$ is used for them
in \cite{8}, but $G(\mathbf{x},\mathbf{x}^\prime)$ is used below
to keep contact with (\ref{38})):
\begin{subequations}
\begin{align}
&G(\mathbf{x},\mathbf{x}^\prime)=
    -\dfrac{\Lambda}{4\pi\beta\hbar^2v^2\rho_{\mathrm{TF}}(S)}\,
\frac1{\mid\mathbf{x}-\mathbf{x}^\prime\mid}\,,\qquad(d=3)\,,
\label{40.1}\\[6pt]
&G(\mathbf{x},\mathbf{x}^\prime)=
    \dfrac{\Lambda}{2\pi\beta\hbar^2v^2\rho_{\mathrm{TF}}(S)}\,
\ln\frac{\mid\mathbf{x}-\mathbf{x}^\prime\mid}{\lambda_T}\,,
     \qquad(d=2)\,,\label{40.2}\\[6pt]
&G(\mathbf{x},\mathbf{x}^\prime)=
  \dfrac\Lambda{2\beta\hbar^2v^2\rho_{\mathrm{TF}}(S)}\,
\mid\mathbf{x}-\mathbf{x}^\prime\mid\,,\qquad\qquad(d=1)\,.
\label{40.3}
\end{align}
\end{subequations}
where $\mathbf{x}$, $\mathbf{x}^\prime$ label spatial arguments at
$d=3$, 2, 1 and $S\equiv\frac12\,(\mathbf{x}+\mathbf{x}^\prime)$.
In (\ref{40.2}), the thermal length $\lambda_T=\hbar\beta v$ is
introduced, where $v=\sqrt{\Lambda/m}$ is the sound velocity given
by (\ref{26}). It is already clear that the correlation functions
can no longer depend on $\mid\mathbf{x}-\mathbf{x}^\prime\mid$
alone: they depend also on the center of mass coordinate $S$,
consistent with the breakdown of translational invariance induced
by the trap.

Let us remind first the correlation functions in $d=3$, 2. In this
case, the points $\mathbf{x}=\mathbf{x}_1$ and
$\mathbf{x}=\mathbf{x}_2$ in
$\varphi_1\equiv\varphi_1(\mathbf{x};\mathbf{x}_1,\mathbf{x}_2)$
are singular and introduce a divergence problem \cite{12}. This
difficulty can be avoided by considering a first-order
\textit{coherence function} $\Gamma^{(1)}({\bf x}_1,{\bf x}_2)$
which is defined in \cite{8} as
\begin{equation}
\Gamma^{(1)}({\bf x}_1,{\bf x}_2)\simeq \frac{\Gamma ({\bf
x}_1,{\bf x}_2)}{\langle\psi_o({\bf
x}_1)\rangle\,\langle\bar\psi_o({\bf x}_2)\rangle}\,,
\label{41}
\end{equation}
where $\Gamma({\bf x}_1,{\bf x}_2)$ is given by (\ref{36}). Then,
$\Gamma^{(1)}({\bf x}_1,{\bf x}_2)$ is both finite and well
defined because identically the same singularities appear
\cite{12} in a direct calculation of $\langle\psi_o({\bf
x}_1)\rangle$ and $\langle\bar\psi_o({\bf x}_2)\rangle $. We find
that
\begin{equation}
\Gamma^{(1)}({\bf x}_1,{\bf x}_2)\simeq
\exp\left(\frac{\Lambda}{4\pi\beta\hbar^2v^2\rho_{\mathrm{TF}}(S)}\,
\frac1{\mid{\bf x}_1-{\bf x}_2\mid}\right)\,.
\label{42}
\end{equation}
Evidently for $d=3$,
$$
\Gamma^{(1)}({\bf x}_1,{\bf x}_2)\,\longrightarrow\,
1+\frac{\Lambda}{4\pi\beta\hbar^2v^2\rho_{\mathrm{TF}}(S)}\,
\frac1{\mid{\bf x}_1-{\bf x}_2\mid}
$$
for $\mid{\bf x}_1-{\bf x}_2\mid\gg
\Lambda/(4\pi\beta\hbar^2v^2\rho_{\mathrm{TF}}(S))$, thus
indicating long-range order and long-range coherence. The
\textit{correlation length} given by
$\Lambda/4\pi\beta\hbar^2v^2\rho_{\mathrm{TF}}(S))$ is therefore a
slowly varying function of ${\bf x}_1$ and ${\bf x}_2$. Notice
that we have assumed ${\bf x}_1$ and ${\bf x}_2$ are not close to
the boundaries.

In the case $d=2$, the correlations decay by a power law for arbitrary
small temperatures,
\begin{equation}
\Gamma^{(1)}({\bf x}_1,{\bf x}_2)\simeq \left(\frac{\lambda_T}
{\mid\mathbf{x}_1-
\mathbf{x}_2\mid}\right)^{\Lambda/(2\pi\beta\hbar^2v^2\rho_{\mathrm{TF}}(S))}.
\label{43}
\end{equation}
The exponent of this power-law is proportional to $T$ so that, at very low
temperatures, correlations may thus prevail over almost macroscopic
distances. At $T=0$ exactly, $\Gamma^{(1)}$ will have a non-zero off-set
due to the presence of a true condensate.

For $d=1$, one can obtain for $G$ instead of (\ref{40.3}) an exact
expression as follows (i.e., the term
$\partial_x\rho_{\mathrm{TF}}\,\partial_x\varphi_1$ in the
governing equations is now accounted for):
\begin{equation}
G({\bf x},{\bf x}^{\prime })\equiv G(x,x^\prime)=
\frac{gR_c}{\beta(2\hbar v)^2}\,\ln
\left[\frac{\bigl(1+|x-x^\prime|/(2R_c)\bigr)^2-
(x+x^\prime)^2/(4R_c^2)}{\bigl(1-|x-x^\prime|/(2R_c)\bigr)^2-
(x+x^\prime)^2/(4R_c^2)}\right]. \label{44}
\end{equation}
Then, we obtain
$\Gamma(\mathbf{x}_1,\mathbf{x}_2)\equiv\Gamma(x_1,x_2)$ in the
following form:
\begin{equation}
\Gamma(x_1,x_2)\simeq
\sqrt{\rho_{\mathrm{TF}}(x_1)\rho_{\mathrm{TF}}(x_2)}
\left[\frac{1+|x_1-x_2|/R_c-x_1x_2/R_c^2}
{1-|x_1-x_2|/R_c-x_1x_2/R_c^2}\right]^{-gR_c/(4\beta\hbar^2v^2)}.
\label{45}
\end{equation}
In the limit $\mid{\bf x}_1-{\bf x}_2\mid\ll\bigl\{R_c,S\bigr\}$,
we obtain $\Gamma(x_1,x_2)$ in the form:
\begin{equation}
\Gamma(x_1,x_2)\simeq
\sqrt{\rho_{\mathrm{TF}}(x_1)\rho_{\mathrm{TF}}(x_2)}\,\,
\exp\left(-\frac{\Lambda}{2\beta\hbar^2v^2\rho_{\mathrm{TF}}(S)}\,|x_1-x_2|\right).
\label{46}
\end{equation}
The correlation length depends now on $x_1$ and $x_2$. Without the
trap, the correlation length reduces to
$2\beta\hbar^2\rho_{\mathrm{TF}}(0)/m$, where
$\rho_{\mathrm{TF}}(0)$ is the density of the ground state of a
homogeneous system at zero temperature, in complete agreement with
the exact solution \cite{6}.

It is obvious that the correlation functions (\ref{43}) and
(\ref{46}) vanish for large separation of the arguments at
arbitrary small temperatures $T>0$, and that there is no
long-range order in $d=2$ or $d=1$. The correlation functions
(\ref{42}), (\ref{43}), (\ref{46}) coincide with those obtained
under translational invariance without the trap to the extent that
there is now an additional factor $\rho_{\mathrm{TF}}(S)$ in the
exponents. True long-range order arises only in $d=3$.

\section{The asymptotics of the correlation functions}

Therefore, the problem concerning the study of the asymptotical
behaviour of the two-point thermal correlation function
$\Gamma(x_1,\tau_1;x_2,\tau_2)$ given by the representation
(\ref{39}), is reduced to solution of the non-homogeneous
Stringari equation (\ref{37}). The corresponding answer (or its
asymptotics) should be subsequently substituted into (\ref{39}).
In the present section, we shall obtain explicitly solutions of
(\ref{37}), and we shall consider the corresponding asymptotics of
$\Gamma(x_1,\tau_1;x_2,\tau_2)$. Let us begin with the limiting
case of a homogeneous Bose gas.

\subsection{The homogeneous Bose gas}

The homogeneous case is given by $V(x)\equiv0$, and the related
equation appears from (\ref{37}) at $1/R_c\to0$:
\begin{equation}
\frac1{\hbar^2v^2}\,\partial_\tau^2G(x,\tau;x^\prime,\tau^\prime)+
\partial_x^2G(x,\tau;x^\prime,\tau^\prime)=
\frac{g}{\hbar^2v^2}\,\delta(x-x^\prime)\delta(\tau-\tau^\prime)\,.
\label{47}
\end{equation}
We consider (\ref{47}) for the domain
$[-R_c,R_c]\times[0,\beta]\ni(x,\tau)$ with the periodic boundary
conditions for each variable. The $\delta$-functions in right-hand
side of (\ref{47}) are treated as the periodic $\delta$-functions.
This allows us to represent the solution of this equation as the
formal double Fourier series:
\begin{equation}
G(x,\tau;x^\prime,\tau^\prime) = \Bigl(\frac{-g}{2\beta
R_c}\Bigr)\,\sum_{\omega,k}
\frac{\E^{\I\omega(\tau-\tau^\prime)+\I k(x-x^\prime)}}
{\omega^2+E_k^2}\,,
\label{48}
\end{equation}
where $\omega=(2\pi/\beta)l$, $l\in\mathbb{Z}$. The notation for
the energy $E_k=\hbar vk$, where $k=(\pi/R_c)n$, $n\in\mathbb{Z}$,
is used in (\ref{48}). Besides, the representation (\ref{48})
requires a regularization, which consists in neglect of the term
given by $\omega=k=0$.

Let us deduce from (\ref{48}) two important asymptotical
representations for the Green function. Then, in the limit of zero
temperature and of infinite size of the domain occupied by the
Bose gas, one can go over to the asymptotics of
$\Gamma(x_1,\tau_1;x_2,\tau_2)$. When a strong inequality
$\beta^{-1}\equiv k_BT\gg\hbar v/R_c$ is valid, we obtain:
\begin{equation}
\begin{array}{l}
G(x,\tau;x^\prime,\tau^\prime)\simeq\\[6pt]
\quad\simeq \dfrac{g}{2\pi\hbar v}\ln\left\{2\Big|
\sinh\dfrac{\pi}{\hbar\beta v}\,\bigl(|x-x^\prime|+ \I\hbar
v(\tau-\tau^\prime)\bigr)\Big|\right\}- \dfrac{g}{4\beta
R_c}\,\dfrac{|x-x^\prime|^2}{\hbar^2v^2}+{\cal C}\,,
\end{array}
\label{49}
\end{equation}
where $|x-x^\prime|\leq 2R_c$, $|\tau-\tau^\prime|\leq \beta$, and
${\cal C}$ is some constant, which is not written explicitly. When
an opposite inequality $\beta^{-1}\equiv k_BT\ll\hbar v/R_c$ is
valid, we obtain:
\begin{equation}
\begin{array}{l}
G(x,\tau;x^\prime,\tau^\prime)\simeq\\[6pt]
\quad\simeq \dfrac{g}{2\pi\hbar v}\ln\left\{2\Big|\sinh
\dfrac{\I\pi}{2R_c}\,\bigl(|x-x^\prime|+\I\hbar
v(\tau-\tau^\prime)\bigr)\Big|\right\}- \dfrac{g}{4\beta
R_c}\,|\tau-\tau^\prime|^2+{\cal C}^\prime\,,
\end{array}
\label{50}
\end{equation}
where $|x-x^\prime|\leq 2R_c$, $|\tau-\tau^\prime|\leq \beta$, and
${\cal C}^\prime$ is another constant.

Let us substitute the estimate (\ref{49}) into the representation
(\ref{39}) and take simultaneously the limit $\beta\hbar v/R_c\to
0$ (the size is growing faster than inverse temperature). Then, we
obtain the following expression for the correlator in question:
\begin{equation}
\Gamma(x_1,\tau_1;x_2,\tau_2)\simeq
\sqrt{\widetilde\rho(x_1)\widetilde\rho(x_2)}\,\,
\Big|\sinh\frac{\pi}{\hbar\beta v}\bigl(|x_1-x_2|+ \I\hbar
v(\tau_1-\tau_2)\bigr)\Big|^{-g/2\pi\hbar v}. \label{51}
\end{equation}
Further, applying the relation (\ref{50}) and taking the limit
$R_c/(\beta\hbar v)\to0$ (the inverse temperature grows faster
than the size), we obtain for $\Gamma(x_1,\tau_1;x_2,\tau_2)$:
\begin{equation}
\Gamma(x_1,\tau_1;x_2,\tau_2)\simeq
\sqrt{\widetilde\rho(x_1)\widetilde\rho(x_2)}\,\,
\Big|\sinh\frac{\I\pi}{2R_c}\,\bigl(|x_1-x_2|+ \I\hbar
v(\tau_1-\tau_2)\bigr)\Big|^{-g/2\pi\hbar v}.
\label{52}
\end{equation}

It follows from (\ref{51}) and (\ref{52}), that in the limit of
zero temperature, $(\hbar\beta v)^{-1}\to 0$, and of infinite
size, $1/R_c\to 0$, the two-point correlation function behaves
like
\begin{equation}
\Gamma(x_1,\tau_1;x_2,\tau_2)\simeq
\frac{\sqrt{\widetilde\rho(x_1)\widetilde\rho(x_2)}}
{\bigl||x_1-x_2|+\I\hbar v(\tau_1-\tau_2)\bigr|^{1/\theta}}\,.
\label{53}
\end{equation}
The latter formula is valid in the limit $\beta\hbar v/R_c\to 0$,
as well as in the limit $R_c/(\beta\hbar v)\to 0$. In (\ref{53})
$\theta$ denotes the critical exponent: $\theta\equiv 2\pi\hbar
v/g$, and the arguments $x_1$ and $x_2$, $\tau_1$ and $\tau_2$ are
assumed to be sufficiently close each to other. Using the
notations $v={\sqrt{\Lambda/m}}$ for the sound velocity and
$\rho=\Lambda/g$ for the density of the homogeneous Bose gas, we
obtain for the critical exponent the following universal
expression:
\begin{equation}
\theta=\frac{2\pi\hbar\rho}{m v}\,.
\label{54}
\end{equation}

\subsection{The trapped Bose gas. High temperature case:
\bmth{k_BT\gg \hbar v/R_c}}

Let us turn to non-homogeneous Bose gas described by equations
(\ref{1})--(\ref{3}). Now, we should consider the non-homogeneous
Stringari equation (\ref{37}) for the arguments
$(x,\tau)\in[-R_c,R_c]\times[0,\beta]$ with the periodic boundary
condition only respectively to $\tau$ (contrary to equation
(\ref{47}), $\delta(x-x^\prime)$ is a usual Dirac's
$\delta$-function supported at the point $x^\prime\in\R$). The
Green function satisfying (\ref{37}) can be written as a formal
Fourier series:
\begin{equation}
G(x,\tau;x^\prime,\tau^\prime)=
\frac1\beta\sum_\omega\E^{\I\omega(\tau-\tau^\prime)}
G_\omega(x,x^\prime)\,,
\label{55}
\end{equation}
where $\omega=(2\pi/\beta)l$, $l\in\mathbb{Z}$. The spectral
density $G_\omega(x,x^\prime)$ in (\ref{55}) is then governed by
the equation
\begin{equation}
-\frac{\omega^2}{\hbar^2v^2}\,G_\omega(x,x^\prime)+ \frac{\D}{\D
x}\left(\Bigl(1-\frac{x^2}{R_c^2}\Bigr) \frac{\D}{\D x}\,G_\omega
(x,x^\prime)\right)= \frac{g}{\hbar^2v^2}\,\delta(x-x^\prime)\,.
\label{56}
\end{equation}

Solution of equation (\ref{56}) can be obtained in terms of the
Legendre functions of the first and second kind, $P_\nu(x/R_c)$
and $Q_\nu(x/R_c)$, \cite{32}, which are linearly independent
solutions of the homogeneous Legendre equation (\ref{27}). As a
result we get \cite{9,10}:
\begin{equation}
G_\omega(x,x^\prime)=\Re G_\omega(x,x^\prime)+\I\Im G_\omega
(x,x^\prime)\,,
\label{57}
\end{equation}
where
\begin{equation}
\begin{array}{rcl}
\Re G_\omega(x,x^\prime)&=&\dfrac{gR_c}{2\hbar^2v^2}\,
\epsilon(x-x^\prime)\bigg\{Q_\nu\Bigl(\dfrac{x}{R_c}\Bigr)
 P_\nu\Bigl(\dfrac{x^\prime}{R_c}\Bigr)-
Q_\nu\Bigl(\dfrac{x^\prime}{R_c}\Bigr)P_\nu\Bigl(\dfrac{x}{R_c}\Bigr)\bigg\},\\[10pt]
\Im G_\omega(x,x^\prime)&=&-\dfrac{gR_c}{2\hbar^2v^2}
\bigg\{\dfrac2\pi\,Q_\nu\Bigl(\dfrac{x}{R_c}\Bigr)
 Q_\nu\Bigl(\dfrac{x^\prime}{R_c}\Bigr)+
\dfrac\pi2\,P_\nu\Bigl(\dfrac{x^\prime}{R_c}\Bigr)
 P_\nu\Bigl(\dfrac{x}{R_c}\Bigr)\bigg\},
\end{array}
\label{58}
\end{equation}
$\nu$ looks as follows:
$$
\nu =-\frac12+\sqrt{\frac14-\Bigl(\frac{R_c}{\hbar
v}\Bigr)^2\omega ^2}\,,
$$
and $\epsilon(x-x^\prime)$ is the sign function
$\epsilon(x)\equiv\sign(x)$. Validity of the solution (\ref{57}),
(\ref{58}) can be verified by direct substitution into (\ref{56}),
where an expression for the Wronskian of two linearly independent
solutions $P_\nu$ and $Q_\nu$ \cite{35} should be used.

The Green function (\ref{57}) can be represented in the form,
which allows to study the corresponding asymptotical behaviour.
When the coordinates $x_1$, $x_2$ are chosen to be far from the
boundary of the trap, $x_1$, $x_2\ll R_c$, but at the same time
the inequalities $|x_1-x_2|\ll\frac12\,(x_1+x_2)$ and
$|x_1-x_2|\ll R_c$ are valid, the corresponding limit should be
called as \textit{quasi-homogeneous}. In the case of strong
inequality $\beta^{-1}=k_BT\gg \hbar v/R_c$, we approximately
obtain for non-zero frequencies: $|\omega|\gg \hbar v/(2R_c)$.
Using the standard asymptotics of the Legendre functions
\cite{32,36}, we determine the behaviour of $G_\omega(x,x^\prime)$
in the quasi-homogeneous limit at large $|\omega|$ as follows:
\begin{equation}
G_\omega(x,x^\prime)\simeq -\frac{\Lambda}{2\hbar
v\rho_{\mathrm{TF}}(S)}\, \frac{\exp\bigl(-(\hbar v)^{-1}
|\omega||x-x^\prime|\bigr)}{|\omega|}\,.
\label{59}
\end{equation}
Here $S$ means a half-sum of the spatial arguments of the
correlator, $S\equiv\frac12\,(x_1+x_2)$, and $v$ is given by
(\ref{26}).

Further, we find that the term $\beta^{-1}G_0(x,x^\prime)$ in
(\ref{55}) is just given (in the quasi-homogeneous limit) by
$G({\bf x},{\bf x}^\prime)$ (\ref{40.3}). Using (\ref{40.3}) and
(\ref{59}) for evaluation of the series (\ref{55}), one obtains
the answer:
\begin{equation}
G(x,\tau;x^\prime,\tau^\prime)\simeq \frac{\Lambda}{2\pi\hbar
v\rho_{\mathrm{TF}}(S)}\ln\left\{2\Big|\sinh\frac{\pi}{\hbar\beta
v}\,\bigl(|x-x^\prime|+\I\hbar v(\tau-\tau^\prime)\bigr)
\Big|\right\}.
\label{60}
\end{equation}
Therefore, the Green function (\ref{39}) takes the following form
at $\beta^{-1}\gg\hbar v/R_c$:
\begin{equation}
\Gamma(x_1,\tau_1;x_2,\tau_2)\simeq
\frac{\sqrt{\widetilde\rho(x_1)\widetilde\rho(x_2)}}
{\Bigl|\,\sinh\dfrac{\pi}{\hbar\beta v}\,\bigl(|x_1-x_2|+ \I\hbar
v(\tau_1-\tau_2)\bigr)\Bigr|^{1/\theta(S)}}\,, \label{61}
\end{equation}
where the critical exponent $\theta(S)$ depends now only on the
half-sum of the coordinates $S$:
\begin{equation}
\theta(S)=\frac{2\pi\hbar\rho_{\mathrm{TF}}(S)}{mv}\,.
\label{62}
\end{equation}
The result (\ref{61}), which is valid for the spatially
non-homogeneous case, is in a correspondence with the estimation
(\ref{51}) obtained above for the homogeneous Bose gas. Therefore,
the expression (\ref{61}) is also concerned with validity of the
condition that the size of the domain occupied by the Bose
condensate grows faster than inverse temperature, i.e., with the
condition $\hbar\beta v/R_c\to 0$.

The relation (\ref{61}) can be simplified for two important
limiting cases. Provided the condition
\begin{equation}
1\ll \frac{|x_1-x_2|}{\hbar\beta v}\ll\frac{R_c}{\hbar\beta v}
\label{63}
\end{equation}
is fulfilled in the quasi-homogeneous case, we obtain from
(\ref{61}) that the correlator decays exponentially:
\begin{equation}
\begin{array}{rcl}
\Gamma(x_1,\tau_1;x_2,\tau_2)&\simeq&
\sqrt{\widetilde\rho(x_1)\widetilde\rho(x_2)}\,\,
\exp\left(-\dfrac{1}{\xi(S)}\bigl||x_1-x_2|+\I\hbar
v(\tau_1-\tau_2)\bigr|\right),\\[6pt]
\xi^{-1}(S)&=&\dfrac{\Lambda}{2\beta\hbar^2v^2\rho_{\mathrm{TF}}(S)}\,.
\end{array}
\label{64}
\end{equation}
The correlation length $\xi(S)$ is defined by the relation
(\ref{64}), which depends now on the half-sum of the coordinates:
\begin{equation}
\xi(S)\equiv \frac{\hbar\beta v}{\pi}\,\theta(S)=
\frac{2\hbar^2\beta\rho_{\mathrm{TF}}(S)}m\,.
\label{65}
\end{equation}

In an opposite limit,
\begin{equation}
\frac{|x_1-x_2|}{\hbar\beta v},\,\frac{|\tau_1-\tau_2|}{\beta}\ll
1\ll \frac{R_c}{\hbar\beta v}\,,
\label{66}
\end{equation}
the asymptotics of $\Gamma(x_1,\tau_1;x_2,\tau_2)$ takes the
following form:
\begin{equation}
\Gamma(x_1,\tau_1;x_2,\tau_2)\simeq
\frac{\sqrt{\widetilde\rho(x_1)\widetilde\rho(x_2)}}
{\bigl||x_1-x_2|+\I\hbar v(\tau_1-\tau_2)\bigr|^{1/\theta(S)}}\,.
\label{67}
\end{equation}
The obtained asymptotics (\ref{67}) is analogous to the estimation
(\ref{53}), which characterizes the spatially homogeneous Bose
gas. But the critical exponent $\theta(S)$ in (\ref{67}) differs
from $\theta$ (\ref{54}), since the latter does not depend on the
spatial coordinates.

\subsection{The trapped Bose gas. Low temperature case:
\bmth{k_BT\ll \hbar v/R_c}}

Let us go over to another case, which also admits investigation of
the asymptotical behaviour of the two-point correlator
$\Gamma(x_1,\tau_1;x_2,\tau_2)$. We begin with the non-homogeneous
Stringari equation (\ref{37}), which can be rewritten in the
following form:
\begin{equation}
\begin{array}{rcl}
\partial_\tau^2G(x,\tau;x^\prime,\tau^\prime)&+&
\dfrac{1}{\alpha^2}\,
\partial_{(x/R_c)}\left(\Bigl(1-\dfrac{x^2}{R_c^2}\Bigr)\partial_{(x/R_c)}
G(x,\tau;x^\prime,\tau^\prime)\right)=\\[9pt]
&=&\dfrac{g}{R_c}\,\delta\Bigl(\sfrac{x-x^\prime}{R_c}\Bigr)
\delta(\tau-\tau^\prime)\,,
\end{array}
\label{68}
\end{equation}
where the notation $\alpha\equiv R_c/(\hbar v)$ is introduced. The
asymptotical behaviour can be investigated in two cases:
$\beta/\alpha\ll1$ (the previous subsection) and
$\beta/\alpha\gg1$ (see below). The functions
$$
\sqrt{n+\frac12}\,P_n\Bigl(\frac{x}{R_c}\Bigr)\,, \quad n\ge 0\,,
$$
where $P_n(x/R_c)$ are the Legendre polynomials, constitute a
complete orthonormal system in the space $L_2\,[-R_c,R_c]$. This
fact allows to obtain the following representation for the Green
function $G(x,\tau; x^\prime,\tau^\prime)$ in the form of the
generalized double Fourier series:
\begin{equation}
G(x,\tau;x^\prime,\tau^\prime)= \Bigl(\frac{-g}{\beta R_c}\Bigr)
\sum_\omega
\sum_{n=0}^{\infty}\frac{n+\frac12}{\omega^2+E_n^2}\,P_n\Bigl(\frac{x}{R_c}\Bigr)
P_n\Bigl(\frac{x^{\prime}}{R_c}\Bigr)\E^{\I\omega(\tau-\tau^\prime)}.
\label{69}
\end{equation}
Summation $\sum\limits_\omega$ in (\ref{69}), as well as in
(\ref{48}) and (\ref{55}), goes over the Bose frequencies, and the
following notation for the energy levels (\ref{28}) is adopted:
\begin{equation}
E_n=\hbar\Omega\sqrt{\frac{n(n+1)}2}=\frac{\sqrt{n(n+1)}}{\alpha}\,.
\label{70}
\end{equation}

After summation over the frequencies and after regularization
consisting in neglect of the term corresponding to zero values of
$\omega$ and $n$, $G(x,\tau;x^\prime,\tau^\prime)$ (\ref{69})
takes the form:
\begin{equation}
\begin{array}{l}
\disty G(x,\tau;x^\prime,\tau^\prime)= \Bigl(\frac{-g}{\beta
R_c}\Bigr)
\biggl[\Bigl(\frac{\beta}{2\pi}\Bigr)^2\sum_{l=1}^\infty
\frac{\cos\bigl(\frac{2\pi\Delta\tau}{\beta}\,l\bigr)}{l^2}\,+\\[10pt]
\disty +\frac{\beta}2\sum_{n=1}^\infty
\frac{n+\sfrac12}{E_n}\,P_n\Bigl(\frac{x}{R_c}\Bigr)
     P_n\Bigl(\frac{x^\prime}{R_c}\Bigr)
\Big(\coth\Bigl(\frac12\,\beta E_n\Bigr)
\cosh(E_n\Delta\tau)-\sinh(E_n\Delta\tau)\Big)\biggr],
\end{array}
\label{71}
\end{equation}
where $\Delta\tau\equiv|\tau-\tau^\prime|$. The representation
(\ref{71}) can be studied in both cases: $\beta/\alpha\ll 1$ and
$\beta/\alpha\gg 1$. For instance, using (\ref{71}) at coinciding
arguments $\tau=\tau^\prime$ to obtain $\Gamma(x_1,\tau;x_2,\tau)$
in the limit $\beta/\alpha\ll 1$, we just obtain \cite{10} the
representations (\ref{45}) or (under the quasi-homogeneity
condition) (\ref{46}).

Let us turn to the case $\beta/\alpha\gg 1$, where $\beta E_n\gg
1$, $\forall n \ge 1$. In other words, let us suppose that $k_B
T\ll E_n$ and, so, $k_B T\ll \hbar\Omega$. Then, one obtains from
(\ref{71}):
\begin{equation}
\begin{array}{l}
G(x,\tau;x^\prime,\tau^\prime)=\dfrac{-g\beta}{4R_c}
\bigg[\Bigl(\dfrac12-\dfrac{\Delta\tau}{\beta}\Bigr)^2-\dfrac1{12}\bigg]-\\[10pt]
\disty\qquad\qquad -\frac{g}{2\hbar v}\sum_{n=1}^\infty
\frac{n+\frac12}{\sqrt{n(n+1)}}\,P_n\Bigl(\frac{x}{R_c}\Bigr)
P_n\Bigl(\frac{x^\prime}{R_c}\Bigr)\,
\exp\Bigl(-\sqrt{n(n+1)}\,\frac{\Delta\tau}{\alpha}\Bigr).
\end{array}
\label{72}
\end{equation}

Notice that a difference between two neighbouring energy levels
(\ref{70}) can be estimated. After some appropriate series
expansions, which are valid at $\forall n> 1$, one obtains:
\begin{equation}
E_{n+1}-E_n\approx\frac{1}{\alpha}\Bigl[1+\frac{1}{8n^2}-
       \frac{1}{4n^3}+\dots\Bigr]\,.
\label{73}
\end{equation}
Equation (\ref{73}) demonstrates that the levels (\ref{70}) are
approximately equidistant provided the inverse powers of $n$ are
neglected in (\ref{73}) for sufficiently large $n>n_0$. In its
turn, the following estimation is also valid:
\begin{equation}
\frac{n+1/2}{\sqrt{n(n+1)}}=1+\frac{1}{8n^2}-\frac{1}{8n^3}+\dots\,.
\label{74}
\end{equation}
It is remarkable that the terms $\propto n^{-1}$ are absent both
in (\ref{73}) and (\ref{74}). Let us remind that leading
asymptotical estimations, which are obtainable with so-called
\textit{logarithmic accuracy}, are important for physical
applications. The problem at hands just admits an estimation with
leading logarithmic accuracy, since inverse powers of $n$ can be
omitted with the same (and good) accuracy in (\ref{73}) and
(\ref{74}) at sufficiently large $n$. Convergency of the series
(\ref{72}) is not affected in this situation, since $\Delta\tau$
is non-zero.

It is known that at sufficiently large $n$, the following
asymptotics for the Legendre polynomials $P_n$ is valid \cite{35}:
\begin{equation}
P_n(\cos\vartheta)= \sqrt{\frac{2}{\pi n\sin\vartheta}}\,\cos\Big[
\bigl(n+\sfrac12\bigr)\vartheta-\frac{\pi}4\Big]+O(n^{-3/2})\,,
\quad 0<\vartheta <\pi\,.
\label{75}
\end{equation}
Let us split the sum over $n$ in (\ref{72}) into two parts:
$\sum\limits_{n=1}^{n=n_0}$ and
$\sum\limits_{n=n_0+1}^{n=\infty}$. Further, let us assume that
$n_0$ is large enough to substitute at $n>n_0$ the Legendre
polynomials $P_n$ by their asymptotical expressions given by
(\ref{75}) (and denoted below as $\bar P_n(\cos\vartheta)$). Then,
using (\ref{73}) and (\ref{74}), we can put
$G(x,\tau;x^\prime,\tau^\prime)$ (\ref{72}) into the following
approximate form \cite{10}:
\begin{equation}
\begin{array}{l}
\disty G(x,\tau;x^\prime,\tau^\prime)\approx\dfrac{-g\beta}{4R_c}
\left[\left(\frac12-\frac{\Delta\tau}{\beta}\right)^2-\frac1{12}\right]-
\frac{g}{2\hbar
v}\,\sum_{n=1}^{n_0}\biggl[\frac{n+\frac12}{\sqrt{n(n+1)}}\,
P_n\Bigl(\frac{x}{R_c}\Bigr)P_n\Bigl(\frac{x^\prime}{R_c}\Bigr)\\[12pt]
\disty\qquad\times
\exp\Bigl(-\sqrt{n(n+1)}\,\frac{\Delta\tau}{\alpha}\Bigr)-
\bar{P}_n\Bigl(\frac{x}{R_c}\Bigr)
\bar{P}_n\Bigl(\frac{x^\prime}{R_c}\Bigr)
\exp\Bigl(-\bigl(n+\sfrac12\bigr)\,\frac{\Delta\tau}{\alpha}\Bigr)\biggr]-\\[12pt]
\disty\qquad\qquad -\frac{g}{2\hbar v}\,\E^{-\Delta\tau/(2\alpha)}
\sum_{n=1}^{\infty}t^n \bar{P}_n\Bigl(\frac{x}{R_c}\Bigr)
\bar{P}_n\Bigl(\frac{x^\prime}{R_c}\Bigr)\,,
\end{array}
\label{76}
\end{equation}
where $t\equiv\exp(-\Delta\tau/\alpha)$, and $n_0$ is the number,
which is fixed (its specific value is forbidden to go to
infinity). Expression (\ref{76}) is valid when $\tau$ and
$\tau^\prime$ are close either to zero or to $\beta$. Besides, we
assume that $\tau\ne\tau^\prime$ in order to keep convergency of
(\ref{76}), while the term omitted can be estimated \cite{10}.

Provided a smallness of $x/R_c$, $x^\prime/R_c$ and
$\Delta\tau/\alpha$ is taken into account, a leading logarithmic
behaviour of the series in (\ref{76}) can be established by the
standard tools \cite{35}. Since in the logarithmic approximation
the first two terms in (\ref{76}) are less important in comparison
to the third one, we write down the leading contribution to the
Green function $G(x,\tau;x^\prime,\tau^\prime)$ in the
quasi-homogeneous limit as follows:
\begin{equation}
G(x,\tau;x^\prime,\tau^\prime)\simeq \frac{-\Lambda}{2\pi\hbar
v}\,\frac{1}{\rho_{\mathrm{TF}}(S)}\,
\ln\frac{R_c}{\bigl||x-x^\prime|+ \I\hbar
v(\tau-\tau^\prime)\bigr|}\,, \label{77}
\end{equation}
where $\rho_{\mathrm{TF}}$ is given by (\ref{23}), $S$ is a
half-sum of $x$ and $x^\prime$, and it is assumed that
\begin{equation}
u_*\equiv \frac{\bigl||x-x^\prime|+\I\hbar
v(\tau-\tau^\prime)\bigr|}{R_c}\ll 1\,.
\label{78}
\end{equation}
Besides, at sufficiently large $n_0$, in our consideration it is
more appropriate to keep $n$ instead of $n+\frac12$ in (\ref{75}).
Expression (\ref{77}) takes place provided the following
conditions of validity of the logarithmic estimation are
respected:
\begin{equation}
1 \ll n_0 < \frac{1}{u_*} \ll\frac{1}{u_*}\ln \frac{1}{u_*}\,.
\label{79}
\end{equation}

A specific value of $n_0$ can be related to the size of the trap
$R_c$: at a restricted range of deviations between the spatial
arguments $x$ and $x^\prime$, increasing of $R_c$ implies
increasing of an upper bound for admissible values of $n_0$.
However, due to (\ref{79}), the estimation obtained for
$G(x,\tau;x^\prime,\tau^\prime)$ (\ref{77}) does not depend
explicitly on a specific choice of $n_0$. In the limit
$1/R_c\to0$, the total coefficient in front of the logarithm in
(\ref{77}) acquires the value $-1/\theta$, where the critical
exponent $\theta$ is defined like in (\ref{53}), (\ref{54}).

Eventually, we obtain the following estimation for the two-point
cor\-re\-la\-tor \break$\Gamma (x_1,\tau_1;x_2,\tau_2)$:
\begin{equation}
\Gamma(x_1,\tau_1;x_2,\tau_2) \simeq
\frac{\sqrt{\widetilde\rho(x_1)\widetilde\rho(x_2)}}
{\bigl||x_1-x_2|+\I\hbar v(\tau_1-\tau_2)\bigr|^{1/\theta(S)}}\,,
\label{80}
\end{equation}
where the notation for the critical exponent $\theta(S)$ is given
by (\ref{62}).

The estimation obtained (\ref{80}), where the critical exponent is
$\theta(S)$ (\ref{62}), constitutes the main result of the present
subsection devoted to the case given by $k_BT\ll \hbar v/R_c$.
From a comparison with the spatially homogeneous Bose gas, one can
see that now the derivation of the estimate (\ref{80}) is just
analogous to a transition from the relation (\ref{52}) to the
final asymptotics (\ref{53}). Then, validity of the corresponding
limiting condition $R_c/(\hbar\beta v)\to 0$ means that the result
(\ref{80}) is also due to the fact that the condensate boundary
$R_c$ increases slower than the inverse temperature.

Therefore, under the different conditions, $k_BT\gg \hbar v/R_c$
and $k_BT\ll \hbar v/R_c$, we demonstrated in the present section
that the behaviour of the correlator in the limit of zero
temperature, $(\hbar\beta v)^{-1}\to 0$, and of infinite size of
the trap, $1/R_c\to 0$, is given by the coinciding estimations
(\ref{67}) and (\ref{80}), i.e. the two-point correlation function
$\Gamma(x_1,\tau_1;x_2,\tau_2)$ has a unique power-law behaviour
in this limit.

\section{Conclusion}

The model considered describes a spatially non-homogeneous weakly
repulsive Bose gas subjected to an external harmonic potential.
The functional integral re\-pre\-sen\-ta\-tion for the two-point
correlation function is estimated by means of the stationary phase
approximation. The main results are obtained for the case when the
size of the domain occupied by the quasi-condensate increases,
while the temperature of the system goes to zero. In the
one-dimensional case, the behaviour of the two-point correlation
function at zero temperature is governed by a power law. However,
in contrast with the case of spatial homogeneity of the Bose gas,
the corresponding critical exponent depends on the same spatial
arguments as the correlator itself. It is just the presence of the
external potential which is responsible for the non-homogeneity of
the critical exponent.

\bigskip
\noindent {\small One of the authors (C. M.) is grateful to the
Organizers of the 8th International Conference ``Path Integrals.
From Quantum Information to Cosmology'' for their warm
hospitality, as well as he is also grateful to P. Jizba and to V.
S. Yarunin for useful comments. A partial support by RFBR, No.
04--01--00825, and by the programme of Russian Academy of Sciences
,,Mathematical Methods in Non-Linear Dynamics'' is acknowledged.}

\end {document}